\documentclass{revtex4}

\usepackage{graphicx}

\begin{document}
\bibliographystyle{revtex}

\setlength{\textheight}{241mm}
\setlength{\textwidth}{170mm}


\def\fig#1{Fig.~(\ref{olnessFig:#1})}
\def\fig#1{Fig.~(\ref{fig:#1})}

\def\figi{
\begin{figure}[t]
\includegraphics[width=5.9cm,angle=0]{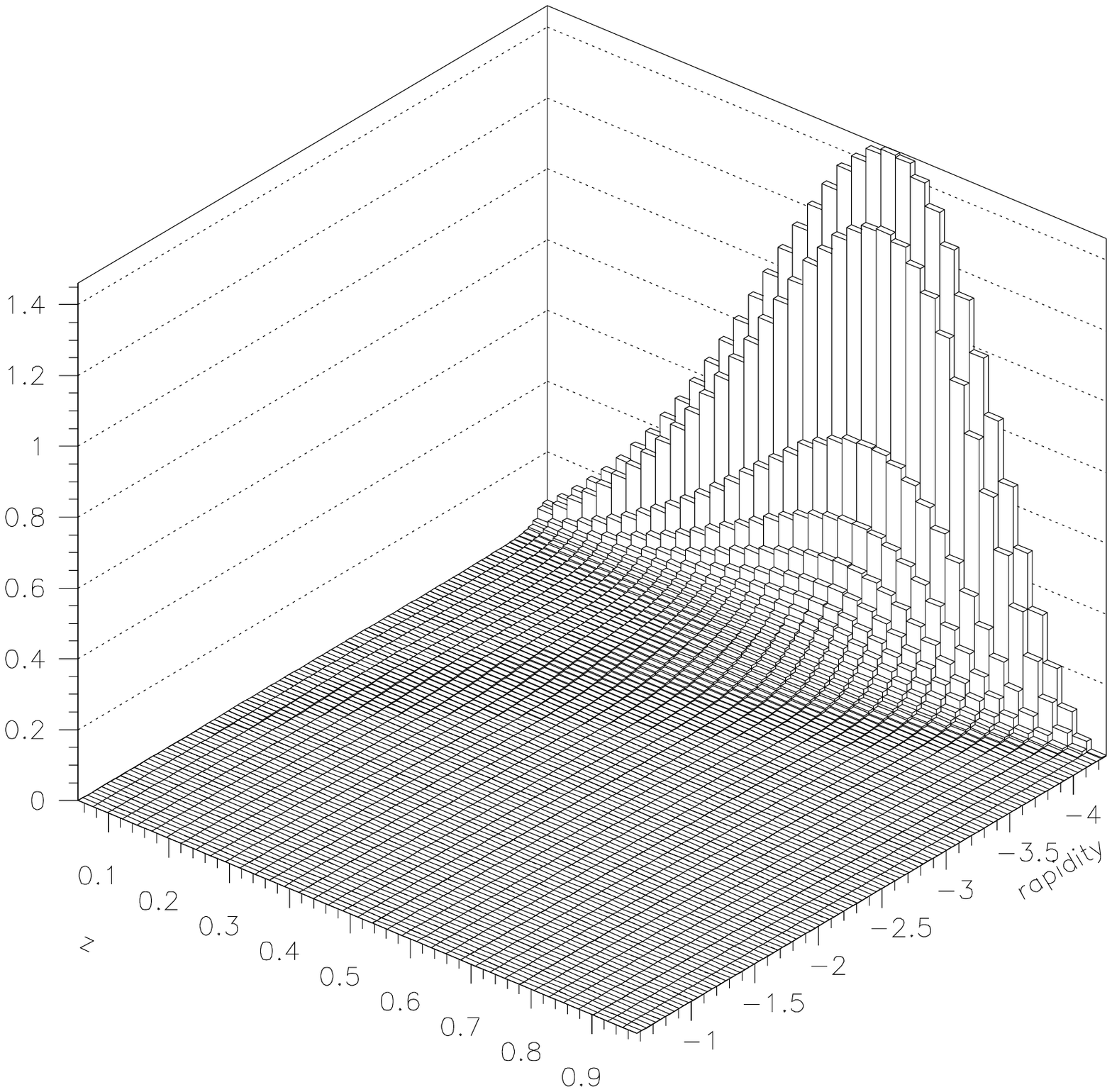}
\includegraphics[width=5.9cm,angle=0]{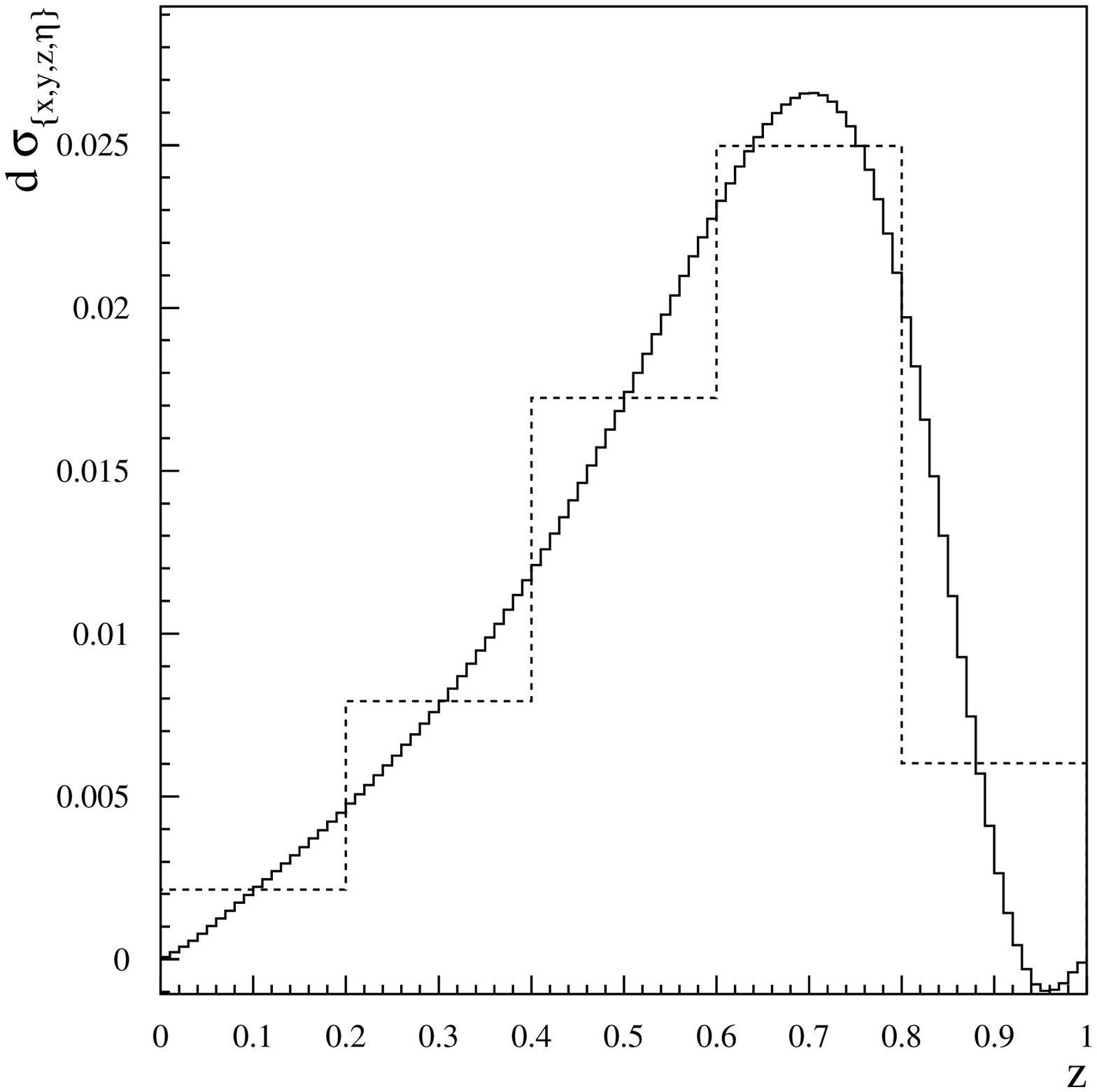}
\includegraphics[width=5.9cm,angle=0]{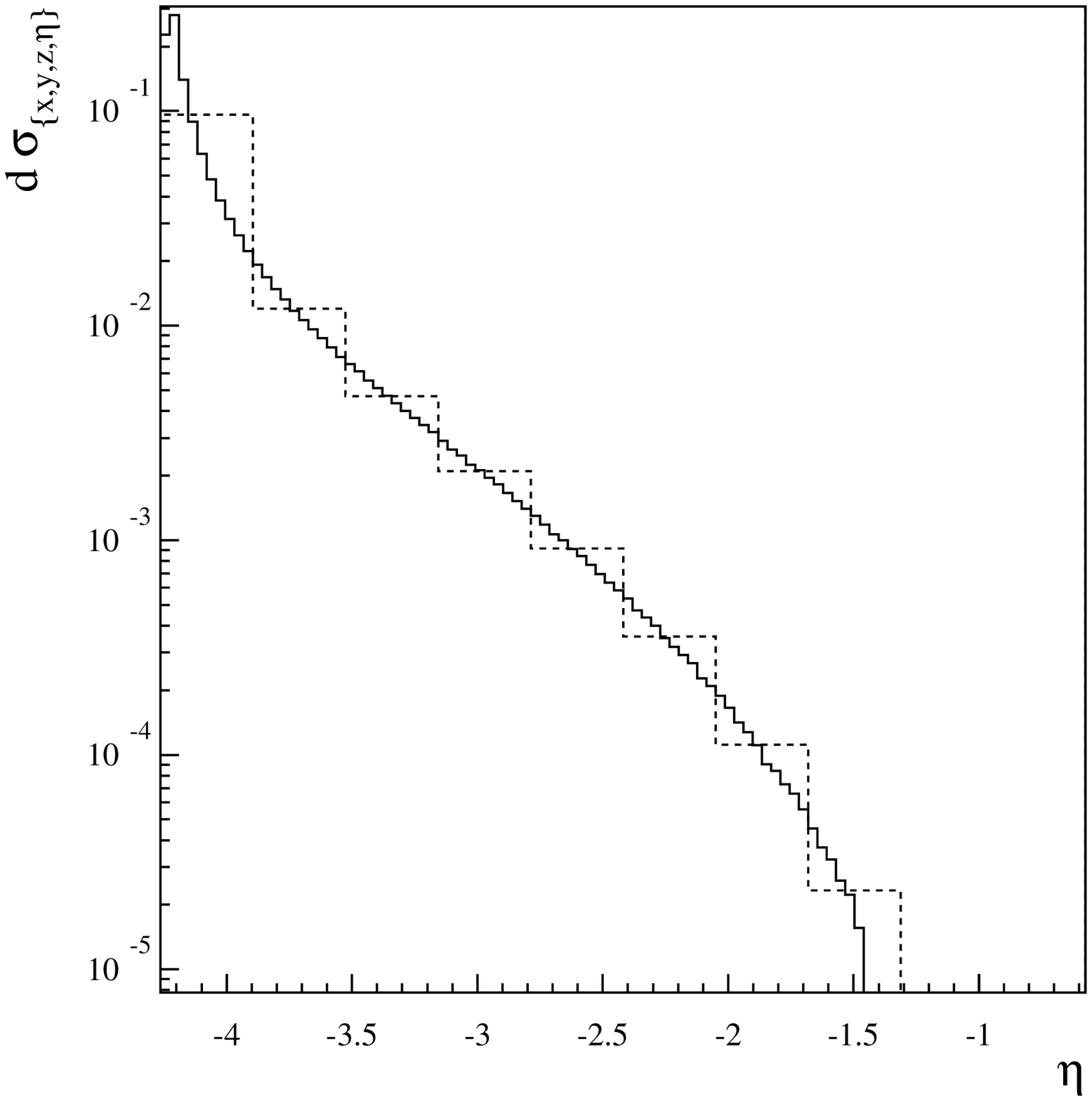}
\hspace*{0.5cm}
\caption{
Differential distribution for charged current 
neutrino-production of charm on an isoscalar target; 
the kinematics shown are for a typical wide-band beam on a fixed
target: $E_\nu = 80\ {\rm GeV}$, $x=0.1$, $Q^2 = 10\ {\rm GeV}^2$.
a)~Binned distribution in both the fragmentation $z$ and 
the charm rapidity $\eta$.
b)~Binned distribution in  $z$ with $\eta$ integrated out.
c)~Binned distribution in  $\eta$ with $z$ integrated out.
Both b) and c) are shown  for a fine and a broad  binning. 
}
\label{fig:olnessFig}
\end{figure}
}

\preprint{hep-ph/0112306}


\title{
Differential Distributions for NLO  Neutrino-Production of Charm}
\thanks{
Contributed to  
 P5-A: Parton Distributions, Spin and Resummation subgroup; 
APS/DPF/DPB Summer Study on the Future of 
Particle Physics (Snowmass 2001), 
Snowmass, Colorado, 30 June - 21 July 2001.
}

\author{S.~Kretzer$^1$, D.~Mason$^2$, F.~Olness$^3$}
\affiliation{
$^1$Department of Physics \&  Astronomy, Michigan State
University, East Lansing, MI 48824 \\
$^2$ Department of Physics, University of Oregon, Eugene, OR 97403\\
$^3$ Department of Physics,  Southern Methodist University, Dallas, TX 75275-0175
}

\date{\today}

\begin{abstract}
\noindent
Charged current DIS charm production measures the strange-quark PDF.
A complete analysis requires both a fully differential theoretical
calculation, and a Monte Carlo detector simulation.  We present
analytic and numeric results at NLO suitable for experimental
analysis.

\end{abstract}

\maketitle

\section{Introduction} \label{sec:intro}

Recent sets of global parton distribution functions (PDFs)
have reached a sufficiently high level of
accuracy that quantization and propagation of statistical errors have
become important issues.  It is, therefore, even more
unsettling that the strange quark PDF, $s(x,Q^2)$, remains a mystery
without a fully consistent picture emerging from the
comparative analysis between neutrino and muon structure functions, 
opposite sign dimuon production in $\nu Fe$-DIS, or the
recently measured parity violating structure function $\Delta xF_3$.  
Given the high precision of the non-strange PDF
components, this situation for $s(x,Q^2)$ is unacceptable both in
terms of our understanding of the nucleon structure, and for our
ability to use precise flavor information to make predictions for
present and future experiments.

For extracting the strange quark PDF, the 
dimuon production data in nu-Fe DIS
provide the most
direct determination.  The basic channel is the weak charged current
process $\nu s \to \mu^- c X$ with a subsequent charm decay $c \to
\mu^+ X^{\prime}$.  These events provide a direct probe of the $s
W$-vertex, and hence the strange quark PDF. In contrast,
single muon production only provides indirect information about
$s(x,Q^2)$ which must then be extracted from a linear combination of
structure functions in the context of the QCD parton model.  For this
reason, fixed-target neutrino dimuon production will provide a unique
perspective on the strange quark distribution of the nucleon in the
foreseeable future.  

In contrast, HERA provides a large dynamic range in
$Q^2$ for the CP conjugated process $e^- {\bar s} \rightarrow {\bar
\nu} c$ which is valuable for testing the underlying QCD evolution.  
Within the HERMES experimental program
the flavor structure of the polarized and unpolarized
sea are studied from semi-inclusive DIS where DIS-Kaon production has
obvious potential to probe strangeness. 
In summary, HERA and HERMES can
complement fixed-target neutrino dimuon data with information at
different energies and from different processes; therefore, neutrino
DIS serves, for now, as an important benchmark process to perform
rigorous and refined comparisons between the experimental data and the
theoretical calculations.  In the long run, a high luminosity neutrino
factory could, of course, considerably raise the accuracy of present
day information from $\nu$-DIS.

The theoretical calculations of inclusive charged current charm
production have been carefully studied in the literature.  Additionally,
the charm fragmentation spectrum has also been calculated in detail.  
While inclusive calculations are sufficient for
many tasks, a comprehensive analysis of the experimental data at NLO
requires additional information from the theoretical side.  In charged
current $\nu$-Fe charm production, the detector acceptance depends on
the full range of kinematic variables: $\{ x, Q^2,z,\eta
\}$.   Here, $x$ is the Bjorken-$x$, $Q$ is the
virtuality of the $W$-boson, $z$ is the scaled energy of the charm
after fragmentation, and $\eta$ is the charm rapidity.  
The theoretical
task is, {\it mutatis mutandis}, not too different from the extraction
of the neutral current charm structure function $F_2^c$ as performed
by the HERA experiments; the HERA analysis uses the
theoretical calculation of the differential cross section
to extrapolate into regions of poor acceptance.

In this short report,\cite{kmo} 
we briefly discuss the key factors that
influence the acceptance of the experimental detector, review the
theoretical calculation of the fully differential cross section at NLO
in QCD, and present numerical results for typical fixed target
kinematics.

\section{Experimental Environment: $\nu$-$Fe$ DIS} \label{sec:exp}

Dimuon events from neutrino charged current charm production can provide a
clear set of events from which to study the strange sea.  Their signiture in a
detector is a pair of oppositely charged muons and hadronic shower originating
from the same vertex.  The second muon is produced in the semileptonic decay of
the charmed particle.  In order to properly reconstruct these events from data,
a minimum energy requirement must be applied to this muon.  To be visible at
all, it must first be energetic enough that it travels further in the detector
than the particles which make up the hadronic shower.  The background from
muons from nonprompt decays of pions and kaons within the shower is large at
low energy, and must also be reduced.  Typically a cut on the charm decay
muon's energy of a few GeV is applied to guarantee it is reconstructable, and
reduce the nonprompt backgrounds.

This energy requirement leads to a dependence on variables in addition to the
typical $E_\nu$, $x$, and $Q^2$ or $y$ dependence of the charged current cross
section.  The energy of the decay muon depends on the energy of the charmed
meson from which it decays, and therefore depends on the fragmentation
parameter $z$.  In NLO, it also depends on the transverse momentum of the
charmed quark.  An event with a low $z$, and/or high transverse momentum will
be less likely to pass a cut on the decay muon's energy than one with a high
$z$ and no transverse momentum.  It is therefore important that the charm
production cross section's dependence on these variables be understood for a
monte carlo to be able to model dimuon event acceptance properly.

\section{Differential Distributions at NLO} \label{sec:theor}

Recorded charged-current charm production rates must be corrected for the
detector acceptance which, as discussed above, depends on the full range of 
kinematic variables  $\{ x, Q^2, z, \eta \}$.
 Therefore, we must obtain the NLO theoretical cross section which is
completely differential in all these variables.
 As the  NLO theoretical cross section contains  
$\delta$-function distributions and ``plus''-distributions, 
there are many inherent difficulties combining this program 
with a complex detector simulation MC program. 
 The singular distributions are 
nothing but an unphysical artifact of
regularized perturbation theory; for any physically observable quantity,
these singularities will be smeared  by soft gluon emission 
to yield physical C-number distributions.

 As the theoretical machinery of soft gluon resummation 
is not fully developed for semi-inclusive DIS (massive) 
heavy quark production, 
we will use a  two step phenomenological approach:
1)~We regularize the NLO calculation  to provide numerical 
distributions free of $\delta$-function and 
``plus''-distributions by integrating over bins
which reflect the finite resolution of the experimental detector.
2)~This result is input to a Monte Carlo (MC) where additional
effects, including iterated soft gluon emissions are added to the 
$p_{\perp}$ smearing
from NLO kinematics to match a Gaussian distribution that has been fit to 
data.

In addition to the above complications, we also encounter large and
negative Sudakov logarithms close to the phase space boundary where,
at fixed-order, soft single-gluon emission is enhanced.  These Sudakov
logarithms diverge in the limit of zero bin-width; 
as we increase our resolution via
narrow binning, we begin to resolve the unphysical $\delta$-functions
and ``plus-distributions.''  Conversely, by using broad bins, we are
effectively integrating over enough phase-space so that the KLN
theorem ensures that we obtain positive physical results.

\section{Numerical Results}\label{sec:num}

\figi

Having addressed the complication of mapping mathematical
distributions onto C-number functions by the introduction of bins, we
now present some preliminary results of step~1) of this calculation.
We compute the normalized differential charm production cross section
for the differential structure function with the binning procedure
defined above.

For our variables, we choose the set 
$\{ x, Q, z, \eta \}$ where $\eta$ is the 
charm rapidity evaluated in the collinear
($p_{\perp ,W}=0$) target rest frame.
 In  \fig{olnessFig}  we present
results for kinematics typical 
of a wide-band neutrino beam on a fixed target:  
$E_\nu = 80\ {\rm GeV}$, $x=0.1$, $Q^2 = 10\ {\rm GeV}^2$. 
 We plot $d \sigma$  in 
2-dimensions {\it vs.} $z$ and $\eta$ in  \fig{olnessFig}-a.
\fig{olnessFig}-b and  \fig{olnessFig}-c show  
$d \sigma_{\{x,y,z,\eta\}}$ for the case where either
$\eta$ or $z$ is integrated out.  In both cases, we display this for
fine-binnings of 1$\times$100 or 100$\times$1, and broad-binnings of
1$\times$5 or 10$\times$1. As an important cross-check on these
results, we verify that the binning $\eta \times z = 1 \times n$
(rapidity integrated out) reproduces the integrated results in the
literature.

We observe negative Sudakov logarithms which occur at large $z$ where
we have $\eta \lesssim \eta_{\max}$ in the integrand.  This effect is
an artifact of our fixed-order result, and is clearly evident in
\fig{olnessFig}-b where we see that in the case of fine binning and large
$z$, the distribution turns negative.
The fine binning effectively resolves 
the unphysical $\delta$-functions and ``plus-distributions,''
and results in an unphysical negative distribution. 
Conversely, by using broad bins, we are effectively integrating over
enough phase-space so that the KLN theorem ensures that we obtain
positive physical results.  In an actual experimental analysis, this
requirement of broad bins arises naturally given the finite detector
resolution. Hence, we observe that negative weights can be easily
avoided by using sufficiently broad bins in a reasonable broadening of
the binning

Our use of bins to regularize the differential distributions will have
negligible impact on the experimental analysis because our bin size is
small compared to the experimental detector resolution, and also
because the detector acceptance is a smooth function in terms of the
set of kinematic variables.  In particular, given the geometry of
typical neutrino detectors, the effective experimental bin size in $z$
and $\eta$ is {\it comfortably} large enough for the purpose of
regularizing the differential distributions with the binning
technique.

\section{Conclusions} \label{sec:conclusions}

In conclusion, 
we have presented  a fully differential NLO calculation of the 
neutrino-induced DIS charm production process. This calculation 
is an essential ingredient for a complete analysis of the 
dimuon data, and will allow a precise determination of the 
strange quark PDF. 
 We have demonstrated that by binning the data appropriately, 
we can interface the theoretical calculation 
(containing $\delta$-functions and ``plus-distributions'')
 directly to the experimental 
Monte Carlo analysis program. 
 We observe the enhancement of the  Sudakov logarithms
at the phase-space boundaries, and verify that these can 
be controlled with this binning method.

The fully differential distributions obtained here 
allow charged current neutrino DIS experiments
to use the complete NLO QCD result in the Monte Carlo 
data analysis. 
 This analysis using the NuTeV data is in progress. 
 These tools will allow us to extract the
strange quark PDF from the dimuon data at NLO with 
unprecedented accuracy; this information
should prove crucial to understanding the behavior of the strange 
quark in the proton.

\section*{Acknowledgment}

The authors would like to thank 
T.~Adams, 
T.~Bolton,
R.~Frey, 
M.~Goncharov,
J.~Morfin, 
R.~Scalise, 
P.~Spentzouris, 
and
W.-K.~Tung, 
for 
helpful discussions.
This research was supported by the
National Science Foundation under Grant PHY-0070443,
by the U.S. Department of Energy,
and by the Lightner-Sams Foundation. 




\end{document}